\documentclass{article}

\usepackage{PRIMEarxiv}

\usepackage[utf8]{inputenc} 
\usepackage[T1]{fontenc}    
\usepackage{microtype}      
\usepackage{amsmath,amssymb,amsthm}
\usepackage{amsfonts}       
\usepackage{nicefrac}       
\usepackage{booktabs}       
\usepackage{tabularx}
\usepackage{array}
\usepackage{subcaption}
\usepackage{graphicx}
\usepackage{xcolor}
\usepackage{listings}
\usepackage{enumitem}
\usepackage{url}            

\usepackage{tikz}
\usetikzlibrary{arrows.meta,positioning,fit}

\definecolor{dvorakGray}{HTML}{333333}
\definecolor{dvorakBlue}{HTML}{003399}
\definecolor{dvorakRed}{HTML}{990000}
\definecolor{dvorakFrame}{HTML}{E0E0E0}

\lstdefinelanguage{lean4}{
  morekeywords=[1]{
    theorem,lemma,def,abbrev,example,axiom,constant,opaque,
    structure,class,instance,inductive,coinductive,
    namespace,section,end,open,import,export,variable,universe,
    where,with,match,if,then,else,let,in,do,return,for,while,
    fun,by,sorry,admit,
    \#print,\#check,\#eval,\#reduce,\#synth
  },
  morekeywords=[2]{
    intro,intros,revert,clear,rename,
    exact,apply,refine,assumption,trivial,rfl,
    have,let,show,suffices,obtain,rcases,cases,induction,match,
    rw,rewrite,simp,simp_all,dsimp,unfold,change,norm_num,ring,linarith,nlinarith,omega,positivity,field_simp,
    constructor,left,right,use,exists,split,ext,funext,congr,
    calc,contradiction,exfalso,by_contra,by_cases,push_neg,contrapose
  },
  morekeywords=[3]{
    Prop,Type,Sort,Nat,Int,Rat,Real,Bool,Char,String,Unit,Empty,
    List,Array,Option,Sum,Prod,Sigma,Subtype,Fin,Set,
    True,False,true,false,
    Event,State,MultitapeTM,KappaBound
  },
  alsoletter={\#_},
  sensitive=true,
  morecomment=[l]{--},
  morecomment=[s]{/-}{-/},
  morestring=[b]",
  literate=
    {ℝ}{{$\mathbb{R}$}}1
    {ℕ}{{$\mathbb{N}$}}1
    {ℤ}{{$\mathbb{Z}$}}1
    {ℚ}{{$\mathbb{Q}$}}1
    {ℂ}{{$\mathbb{C}$}}1
    {≤}{{$\le$}}1
    {≥}{{$\ge$}}1
    {≠}{{$\ne$}}1
    {→}{{$\to$}}1
    {←}{{$\leftarrow$}}1
    {↔}{{$\leftrightarrow$}}1
    {∀}{{$\forall$}}1
    {∃}{{$\exists$}}1
    {∧}{{$\land$}}1
    {∨}{{$\lor$}}1
    {¬}{{$\neg$}}1
    {∈}{{$\in$}}1
    {∉}{{$\notin$}}1
    {⊆}{{$\subseteq$}}1
    {∪}{{$\cup$}}1
    {∩}{{$\cap$}}1
    {∅}{{$\emptyset$}}1
    {×}{{$\times$}}1
    {∘}{{$\circ$}}1
    {⟨}{{$\langle$}}1
    {⟩}{{$\rangle$}}1
    {λ}{{$\lambda$}}1
    {α}{{$\alpha$}}1
    {β}{{$\beta$}}1
    {γ}{{$\gamma$}}1
    {ε}{{$\varepsilon$}}1
    {κ}{{$\kappa$}}1
    {⊢}{{$\vdash$}}1
    {⊨}{{$\models$}}1
}

\lstnewenvironment{lean}{
  \lstset{
    language=lean4,
    basicstyle=\ttfamily\small\color{dvorakGray},
    keywordstyle=[1]\bfseries\color{dvorakRed},
    keywordstyle=[2]\color{dvorakBlue},
    keywordstyle=[3]\bfseries\color{dvorakBlue},
    stringstyle=\color{dvorakRed!80!black},
    commentstyle=\itshape\color{gray!75!black},
    frame=single,
    framerule=0.5pt,
    rulecolor=\color{dvorakFrame},
    xleftmargin=0.8em,
    framexleftmargin=0.4em,
    breaklines=true,
    columns=fullflexible,
    keepspaces=true,
    mathescape=true
  }
}{}

\usepackage[colorlinks=true,linkcolor=blue!55!black,citecolor=blue!55!black,urlcolor=blue!55!black]{hyperref}

\title{ProofVMM: A Maturity Model for Auditing Lean Proofs}

\author{
  Gustas Mickus \\
  \normalfont\small Faculty of Mathematics and Informatics \\
  \normalfont\small Vilnius University \\
  \normalfont\texttt{gustas.mickus@mif.stud.vu.lt}
  \And
  Linas Petkevi\v{c}ius \\
  \normalfont\small Institute of Computer Science \\
  \normalfont\small Vilnius University \\
  \normalfont\texttt{linas.petkevicius@mif.vu.lt}
}

\begin{document}
\maketitle

\begin{abstract}
Interactive theorem provers such as Lean~4 verify whether a proof term inhabits its formal type, yet they do not check whether that type matches the intended source claim or whether auxiliary proofs advance a target theorem. We introduce \textbf{ProofVMM} (the Proof Verification Maturity Model), a five-tier framework ($L_0$--$L_4$) with four transition gates that separates automated verification status ($L_{\mathrm{tech}}$) from audited semantic faithfulness ($L_{\mathrm{cred}}$), defining the \emph{Compile--Faithfulness Gap} as $\Delta = L_{\mathrm{tech}} - L_{\mathrm{cred}}$, and that propagates maturity across a dependency graph. Applying ProofVMM to the 10-mission \emph{Prove2Me} campaign (aiming to multiply integers faster than $O(n \log n)$ time) reveals that whilst 5 missions display a \texttt{Proved} status with zero open leaves, only 3 establish full-scope claims while the other 2 verify fixed-constant special cases. An interactive session with \texttt{DeepSeek V4.1 Flash} on an open frontier target further produces 21 kernel-checked auxiliary theorems that leave the target leaf untouched, exposing totalisation traps, mis-indexed specifications, and scope mismatches. ProofVMM turns binary \texttt{Proved} labels into an auditable breakdown of what has actually been established.
\end{abstract}


\section{Introduction}
\label{sec:intro}

AI agents can now generate thousands of lines of Lean~4 code that pass the type checker, yet ``Lean verified'' has become an ambiguous label. A proof assistant guarantees that a term inhabits its formal type, never that the formal type captures the intended mathematics or advances an open conjecture. Bridging this gap requires more than stronger provers: it demands an explicit lifecycle discipline for proof maturity.

Staged maturity frameworks have repeatedly turned fragile technical workflows into reliable engineering. NASA's nine-level Technology Readiness Levels (TRLs) \cite{mankins1995trl} separated laboratory concepts from flight-qualified systems; CMMI process gates yielded self-reported median improvements of 48\% in product quality and 50\% in schedule across 35~software organisations \cite{gibson2006cmmi}; and MLOps maturity tiers (Levels~0--2) \cite{google2023mlops} bridged manual notebook scripts and reproducible production pipelines. As automated formalisation in proof assistants such as Lean~4 \cite{demoura21lean4} scales from isolated contest exercises to multi-contributor research repositories, formalised mathematics faces an analogous ``LeanOps'' transition. Recent audits show, however, that binary metrics (whether a Lean file compiles without \texttt{sorry}, or how many declarations are marked \texttt{Proved}) give a misleading picture of progress. A Lean declaration can compile cleanly whilst altering hypotheses or bounds \cite{zhang2026compilationevaluatingfaithfulnaturallanguagetolean,bastounis2026navierstokeslosttranslationlean}, and a repository can accumulate valid auxiliary lemmas without discharging its central conjecture \cite{venturim2026leantrail}.

Existing research (ranging from statement-level faithfulness benchmarks \cite{zhang2026compilationevaluatingfaithfulnaturallanguagetolean,amster2026claimcheck} and single-paper alignment tools \cite{venturim2026leantrail} to modified proof-relevant kernels \cite{will2026eigeniustypedknowledgegraphdbms}) addresses isolated parts of this problem (reviewed in Section~\ref{sec:related}). To our knowledge, none provides an end-to-end maturity ladder for unmodified Lean~4 codebases that separates proof closure from statement-closure auditing, incorporates \texttt{.olean} serialisation and \texttt{Mathlib} integration, and evaluates a live multi-mission collaborative platform empirically.

The main three contributions:
\begin{itemize}
    \item \textbf{The ProofVMM Framework (Section~\ref{sec:vmm}).} We define five artefact-based tiers ($L_0$--$L_4$), four transition gates, and a clean separation between proof closure ($L_{\mathrm{tech}}$) and statement-closure credit ($L_{\mathrm{cred}}$).
    \item \textbf{Exhaustive Campaign Audit (Section~\ref{sec:macro_scope}).} Auditing all 10 missions (415~node entries; 289~entries after removing the shared 126-entry Round~7/8 pool) of the \emph{Prove2Me} \texttt{integer-\allowbreak multiplication-\allowbreak kappa} campaign \cite{prove2me2026,beyondnlogn}, we find that 7 of 10 missions fail to reach $L_{\mathrm{cred}} = 3$ (four at $L_{\mathrm{cred}} = 2$ and three at $L_{\mathrm{cred}} = 0$) even though 87.9\% of the 289~entries are marked \texttt{Proved}.
    \item \textbf{Interactive Trace Diagnostic and Error Taxonomy (Sections~\ref{sec:macro_pr6453}--\ref{sec:vmm_taxonomy}).} Examining an interactive \texttt{DeepSeek V4.1 Flash} session on PR~\#6453, we document 21 kernel-checked theorems outside the target's critical path (alongside one RAM-verified instance failing \texttt{.olean} export) and classify three empirical error modes observed in the audit.
\end{itemize}

\section{ProofVMM: The Proof Verification Maturity Model}
\label{sec:vmm}

Treating formal verification as a binary property (``verified'' versus ``unverified'') conflates syntactic type-checking with mathematical truth. To evaluate Lean~4 formalisations systematically at the level of individual claims, we define \textbf{ProofVMM}.

\subsection{Four Principles and Five Maturity Tiers}
\label{sec:vmm_levels}

Four principles govern ProofVMM:
\begin{enumerate}
    \item \textbf{Claim-level granularity.} Each lemma, proposition, or theorem is graded individually rather than at the paper or repository level.
    \item \textbf{Artefact-defined tiers.} Under the Curry--Howard correspondence (propositions as types, proofs as terms), tiers are defined by inspectable properties of the formal artefact.
    \item \textbf{Fidelity separation.} Each claim~$v$ carries two scores in $\{0,\dots,4\}$: a \emph{technical level} $L_{\mathrm{tech}}(v)$, measuring mechanical proof closure, and a \emph{credited level} $L_{\mathrm{cred}}(v)$, granted only after auditing the claim's statement closure against its source claim~$\varphi_S(v)$. Their difference defines the \emph{Fidelity Deficit} (the \emph{Compile--Faithfulness Gap}):
    \begin{equation}
        \Delta(v) \;=\; L_{\mathrm{tech}}(v) - L_{\mathrm{cred}}(v).
        \label{eq:fidelity_deficit}
    \end{equation}
    For open stronger targets unsupported by a weaker conditional source (starred rows in Section~\ref{sec:macro_audit}), $\Delta$ records this provenance gap rather than an unfaithful proof.
    \item \textbf{The Weakest Link Rule.} Technical proof maturity cannot exceed the weakest proof dependency (except that closing a proof locally over audited incomplete components reaches Level~2), whilst credited maturity requires faithful statement closure.
\end{enumerate}

ProofVMM defines five ordered tiers. As a mnemonic (not an isomorphism), the tiers resemble stages of the software lifecycle, indicated in italics.\footnote{A kernel-checked proof covers every instance of its statement, whereas a test suite samples behaviours, and a Lean \texttt{sorry} is a stub flagged by the build that simulates nothing at runtime; the tiers rest strictly on Curry--Howard.}

\begin{itemize}
    \item \textbf{Level 0 (Unverified Prose) --- \emph{Requirements}:} Informal natural-language or \LaTeX\ claim and proof.
    \item \textbf{Level 1 (Syntactic Scaffolding) --- \emph{API Design}:} A well-typed Lean~4 declaration compiling under \texttt{lake build} that ends in \texttt{sorry} or relies on user-declared axioms (e.g.\ \emph{AnnalsChallenge} \cite{browning2026datasetformalisedtheoremstatements}).
    \item \textbf{Level 2 (Staged Blueprint or Incomplete Closure) --- \emph{Integration Mocks and Stubs}:} The declaration's own body has zero \texttt{sorry} and no custom axioms, but relies on open $L_1$ stubs within an acyclic dependency graph, uses reduction-bypassing tactics such as \texttt{native\_decide}, or passes kernel checking in RAM whilst failing \texttt{.olean} export within the evaluation resource budget. (Fixed-constant scoped formalisations can pass Level~3 technically yet remain capped at $L_{\mathrm{cred}} \le 2$ under Convention~(\ref{conv:scoped}).)
    \item \textbf{Level 3 (Isolated Kernel Verification) --- \emph{Unit-Tested Build}:} Complete elimination of \texttt{sorry} placeholders and custom axioms across the declaration and all project dependencies. Executing \texttt{\#print axioms} outputs strictly standard foundational axioms (\texttt{propext}, \texttt{Classical.choice}, \texttt{Quot.sound}), forbidding \texttt{Lean.ofReduceBool} (from \texttt{native\_decide}) and \texttt{unsafe} code. The build and binary \texttt{.olean} export are reproducible on a pinned toolchain within a declared resource budget (in our case study, the default SAIR OpenMath~OS container memory limit; see Section~\ref{sec:macro_pr6453}).
    \item \textbf{Level 4 (Ecosystem Integration) --- \emph{Production Release}:} Full integration adhering to \texttt{Mathlib} \cite{mathlib2020} architectural standards: refactored onto canonical typeclass hierarchies, passing repository linters (\texttt{\#lint}) with zero concept duplication, and corroborated by human maintainer Pull Request review.
\end{itemize}

\paragraph{Human-in-the-loop at Level~4.} Achieving Level~4 cannot be fully automated. Whilst scripts can verify compilation and run \texttt{\#lint}, automated formalisations can duplicate abstractions or produce isolated signatures disconnected from the rest of the library. \texttt{Mathlib} maintainers possess deep expertise in the library's typeclass hierarchy, mathematical generality, and design conventions; their \emph{human-in-the-loop} feedback ensures that formalised theorems remain faithful, reusable, and maintainable across the ecosystem.

\subsection{Transition Gates and Semantic Direction}
\label{sec:vmm_gates}

Advancement between tiers is governed by four \emph{Transition Gates} ($G_{0\to 1}$ to $G_{3\to 4}$), summarised in Table~\ref{tab:vmm_gates}; failing a gate leaves the claim at the preceding tier. At Gate~$1\to 2$, deterministic tools such as \texttt{comparator} and \texttt{con-leche} \cite{LFM26,Sut26} can automatically verify that a declaration and its unfolded definitions match an existing trusted Lean reference statement, whilst automated LLM judges \cite{zhang2026compilationevaluatingfaithfulnaturallanguagetolean,amster2026claimcheck,venturim2026leantrail} may pre-screen natural-language correspondence if calibrated against human labels. Still, human confirmation against the source manuscript remains essential: \emph{LeanTrail} reports that 17.9\% of \texttt{Gemini 3.8 Flash}'s \texttt{Formalized} verdicts were false positives on weakened statements, whilst \texttt{GPT-6 Sol} never over-claimed yet recalled only 52.4\% of formalised items \cite{venturim2026leantrail}. For constructive results, explicit witnesses should be defined rather than bypassed via non-constructive existence theorems \cite{yang2026constructionverificationbenchmarkappliedmathematics}.

\begin{table}[htbp]
\centering
\small
\caption{Summary of ProofVMM Transition Gates ($G_{0\to 1}$ to $G_{3\to 4}$).}
\label{tab:vmm_gates}
\begin{tabularx}{\textwidth}{@{}l >{\raggedright\arraybackslash}X >{\raggedright\arraybackslash}p{3.2cm} >{\raggedright\arraybackslash}X@{}}
\toprule
\textbf{Gate} & \textbf{Automatic Checks} & \textbf{Manual Audit} & \textbf{Not Guaranteed} \\
\midrule
$G_{0\to 1}$ & \texttt{lake build} succeeds; signature type-checks & Claim card completed & Semantic fidelity \\
$G_{1\to 2}$ & Own body has zero \texttt{sorry} and no user \texttt{axiom}; \texttt{comparator}/\texttt{con-leche} against Lean reference (if available) & Source--statement closure audit & Truth of stubbed lemmas \\
$G_{2\to 3}$ & Dependencies complete; standard axioms only; no \texttt{native\_decide}; \texttt{.olean} built within budget & --- & Fidelity; constructivity \\
$G_{3\to 4}$ & Current \texttt{Mathlib} build; \texttt{\#lint} passes & Zero duplication; maintainer PR review & Future \texttt{Mathlib} changes \\
\bottomrule
\end{tabularx}
\end{table}

\paragraph{Semantic direction.} At Gate~$1\to 2$, a formal statement is compared against its source claim by inspecting its premises and conclusion over shared domains. Adding an extra hypothesis or narrowing the conclusion makes the Lean statement \emph{weaker}: proving it does not establish the source claim ($L_{\mathrm{cred}} = 0$, unless recorded as a scoped instantiation under Convention~(\ref{conv:scoped})). Altering the domain (for instance through default totalisations) renders the formal statement \emph{incomparable} ($L_{\mathrm{cred}} = 0$). Conversely, dropping assumptions from a conditional source makes the Lean target \emph{stronger}: proving a stronger target at $L_{\mathrm{tech}} \ge 3$ \emph{would} establish the source claim, yet whilst the target remains open ($L_{\mathrm{tech}} < 3$), the weaker conditional source evidence cannot support it (tagged as an \emph{unsupported stronger target} with $L_{\mathrm{cred}} = 0$ relative to the conditional source provenance).

\paragraph{Operational Conventions.}
\label{sec:vmm_conventions}

Three conventions resolve boundary cases during grading and allow direct cross-referencing (\ref{conv:deps}--\ref{conv:external}):

\begin{enumerate}[label=\textbf{(\roman*)}, ref=\roman*]
    \item \label{conv:deps} \textbf{Dependencies versus Assumptions.}
    If a source claim has the conditional form ``if $H$ then $C$'', the hypothesis~$H$ belongs to the statement's type rather than to open proof dependencies: a closed proof of $H \to C$ reaches~$L_3$, subject to the semantic direction criteria above.
    \emph{Example:} Proving an optimisation bound under an explicit convexity assumption is a valid $L_3$ result. However, silently replacing convexity with strong convexity weakens the claim and fails Gate~$1\to 2$ ($L_{\mathrm{cred}} = 0$).

    \item \label{conv:scoped} \textbf{Scoped versus Full Claims.}
    When a formalisation proves only a fixed-constant instantiation or special case of a general parametric claim, its credited level is capped at $L_{\mathrm{cred}} \le 2$, with the missing generality recorded as a \emph{scoped gap}.
    \emph{Example:} Proving a bound only for dimension $d = 2$ or $2 \times 2$ matrices reaches $L_{\mathrm{tech}} = 3$ locally, while the general claim for arbitrary~$d \in \mathbb{N}$ remains capped at $L_{\mathrm{cred}} \le 2$.

    \item \label{conv:external} \textbf{Kernel Verification versus External Evidence.}
    Only proof terms verified by the Lean~4 kernel raise $L_{\mathrm{tech}}$ or $L_{\mathrm{cred}}$; external scripts and numerical certificates are logged separately as auxiliary evidence.
    \emph{Example:} Validating a bound via Monte Carlo simulations in Python or checking a 60-digit inequality via \texttt{mpmath} builds empirical confidence, yet leaves $L_{\mathrm{tech}} = 0$ until formalised in Lean.
\end{enumerate}

\subsection{Dependency Propagation: Proof Closure vs.\ Statement Closure}
\label{sec:vmm-principles}

A project is represented as a finite directed acyclic graph (DAG) $\mathcal{G} = (V, E)$, where $u \to v$ means claim~$v$ references~$u$ in its definition or proof. Let $\operatorname{deps}(v) = \{u \in V \mid (u, v) \in E\}$ denote the direct dependencies of~$v$. Core Lean and \texttt{Mathlib} declarations are trusted.

To avoid conflating proof verification with statement fidelity, ProofVMM evaluates technical proof closure and credited statement maturity via two distinct propagation rules:

\paragraph{Technical level (proof closure).}
A claim's mechanical proof maturity is bounded by its dependencies:
\begin{equation}
L_{\mathrm{tech}}(v) \;=\; \min\Big(L_{\mathrm{tech}}^{\mathrm{loc}}(v),\ \min_{u \in \operatorname{deps}(v)} L_{\mathrm{tech}}(u)\Big),
\label{eq:weakest_link}
\end{equation}
except that closing a proof locally ($L_{\mathrm{tech}}^{\mathrm{loc}}(v) \ge 2$) over open $L_1$ stubs counts as Level~2 (a staged blueprint). Here $L_{\mathrm{tech}}^{\mathrm{loc}}(v)$ reflects $v$'s own proof body: 0 if unelaborated, 1 if containing \texttt{sorry} or user axioms, 2 if using \texttt{native\_decide} or failing \texttt{.olean} export, 3 if kernel-verified and serialisable, and 4 if meeting \texttt{Mathlib} linter and style standards.

\paragraph{Credited level (statement closure).}
Semantic fidelity is evaluated on the claim's \emph{statement closure}: the fully elaborated type of~$v$ and the non-trusted definitions, notations, and coercions it unfolds. Auditing against the source claim~$\varphi_S(v)$ assigns a credit cap (calculated as $\mathrm{cap}(v) \in \{0, 2, 4\}$: 4 for faithful statements, 2 for fixed-constant scoped claims under Convention~(\ref{conv:scoped}), and 0 for weaker, incomparable, or unsupported open targets). The credited level is:
\begin{equation}
L_{\mathrm{cred}}(v) \;=\; \min\Big(L_{\mathrm{tech}}(v),\ \mathrm{cap}(v),\ \min_{u \in \operatorname{open\_stubs}(v)} \mathrm{cap}(u)\Big),
\label{eq:lcred}
\end{equation}
where $\operatorname{open\_stubs}(v)$ is the set of direct or transitive dependencies of~$v$ that are not yet kernel-certified ($L_{\mathrm{tech}} < 3$, with $\min \emptyset = 4$).

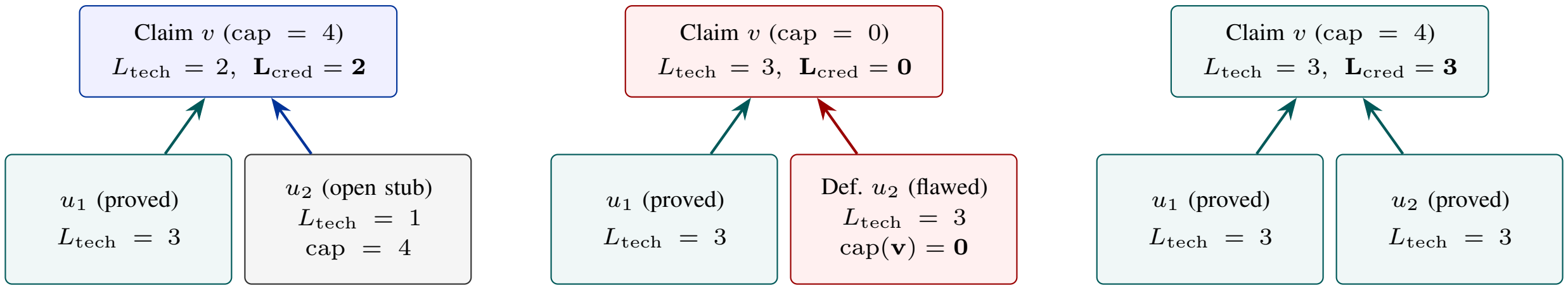
\begin{figure}[htbp]
\centering
\begin{subfigure}[t]{0.31\textwidth}
\centering
\begin{tikzpicture}[>=Stealth, every node/.style={rounded corners=2pt, align=center, font=\scriptsize, inner sep=3pt}]
  \node[draw=teal!70!black, fill=teal!6, text width=2.15cm, minimum height=1.35cm] (u1) at (-1.25, 0)
    {$u_1$ (proved)\\[3pt] $L_{\mathrm{tech}}=3$};
  \node[draw=dvorakGray, fill=gray!8, text width=2.15cm, minimum height=1.35cm] (u2) at (1.25, 0)
    {$u_2$ (open stub)\\[1pt] $L_{\mathrm{tech}}=1$\\[1pt] $\mathrm{cap}=4$};
  \node[draw=dvorakBlue, fill=blue!6, text width=3.1cm, minimum height=0.95cm] (v) at (0, 1.75)
    {Claim $v$ ($\mathrm{cap}=4$)\\[2pt] $L_{\mathrm{tech}}=2,\; \mathbf{L_{\mathrm{cred}}=2}$};
  \draw[->, thick, teal!70!black] (u1) -- (v);
  \draw[->, thick, dvorakBlue] (u2) -- (v);
\end{tikzpicture}
\caption{Conditional Closure ($L_2$): audited open stub~$u_2$ caps $v$ at Level~2.}
\end{subfigure}
\hfill
\begin{subfigure}[t]{0.31\textwidth}
\centering
\begin{tikzpicture}[>=Stealth, every node/.style={rounded corners=2pt, align=center, font=\scriptsize, inner sep=3pt}]
  \node[draw=teal!70!black, fill=teal!6, text width=2.15cm, minimum height=1.35cm] (u1) at (-1.25, 0)
    {$u_1$ (proved)\\[3pt] $L_{\mathrm{tech}}=3$};
  \node[draw=dvorakRed, fill=red!6, text width=2.15cm, minimum height=1.35cm] (u2) at (1.25, 0)
    {Def.~$u_2$ (flawed)\\[1pt] $L_{\mathrm{tech}}=3$\\[1pt] $\mathbf{\mathrm{cap}(v)=0}$};
  \node[draw=dvorakRed, fill=red!6, text width=3.1cm, minimum height=0.95cm] (v) at (0, 1.75)
    {Claim $v$ ($\mathrm{cap}=0$)\\[2pt] $L_{\mathrm{tech}}=3,\; \mathbf{L_{\mathrm{cred}}=0}$};
  \draw[->, thick, teal!70!black] (u1) -- (v);
  \draw[->, thick, dvorakRed] (u2) -- (v);
\end{tikzpicture}
\caption{Statement Failure: flawed definition~$u_2$ inside $v$'s statement yields $\Delta = +3$.}
\end{subfigure}
\hfill
\begin{subfigure}[t]{0.31\textwidth}
\centering
\begin{tikzpicture}[>=Stealth, every node/.style={rounded corners=2pt, align=center, font=\scriptsize, inner sep=3pt}]
  \node[draw=teal!70!black, fill=teal!6, text width=2.15cm, minimum height=1.35cm] (u1) at (-1.25, 0)
    {$u_1$ (proved)\\[3pt] $L_{\mathrm{tech}}=3$};
  \node[draw=teal!70!black, fill=teal!6, text width=2.15cm, minimum height=1.35cm] (u2) at (1.25, 0)
    {$u_2$ (proved)\\[3pt] $L_{\mathrm{tech}}=3$};
  \node[draw=teal!70!black, fill=teal!6, text width=3.1cm, minimum height=0.95cm] (v) at (0, 1.75)
    {Claim $v$ ($\mathrm{cap}=4$)\\[2pt] $L_{\mathrm{tech}}=3,\; \mathbf{L_{\mathrm{cred}}=3}$};
  \draw[->, thick, teal!70!black] (u1) -- (v);
  \draw[->, thick, teal!70!black] (u2) -- (v);
\end{tikzpicture}
\caption{Kernel Verified ($L_3$): statement closure faithful; proof closed at $L_3$.}
\end{subfigure}
\caption{Proof closure ($L_{\mathrm{tech}}$, Eq.~\eqref{eq:weakest_link}) versus statement-closure credit ($L_{\mathrm{cred}}$, Eq.~\eqref{eq:lcred}).}
\label{fig:weakest_link}
\end{figure}

Under this rule, proved internal helper lemmas ($L_{\mathrm{tech}} \ge 3$) require no statement audit: if the Lean kernel accepts the complete proof term, a bizarre or trivial helper (such as using \texttt{Real.log (x / x) = 0}) cannot falsify a faithful root statement. Only the statement closure of~$v$ and any unfinished open stubs can cap credit (Figure~\ref{fig:weakest_link}).

\subsection{Worked Example: Step-by-Step Progression on AM--GM}
\label{sec:vmm_worked}

To show how a claim advances across the ProofVMM tiers and gates, we trace the elementary arithmetic mean--geometric mean inequality:
\begin{equation*}
    \forall a, b \in \mathbb{R},\quad a \ge 0 \;\land\; b \ge 0 \implies \sqrt{ab} \le \frac{a+b}{2}.
\end{equation*}

\paragraph{Level 0: Unverified Prose.}
On paper, the claim is stated in natural language or \LaTeX\ and proved informally by expanding $(\sqrt{a} - \sqrt{b})^2 = a - 2\sqrt{ab} + b \ge 0$. A \emph{claim card} records the requirements: two real variables $a, b \in \mathbb{R}$, hypotheses $a \ge 0$ and $b \ge 0$, and the target inequality. With no formal artefact, $L_{\mathrm{tech}} = 0$ and $L_{\mathrm{cred}} = 0$.

\paragraph{Level 1: Syntactic Scaffolding and Semantic Drift.}
Passing Gate~$0\to 1$ requires a Lean~4 signature that compiles under \texttt{lake build} with \texttt{sorry}:
\begin{lean}
-- Faithful L1 declaration (passes G$_{0\to 1}$, audited: L_tech = 1, cap = 4 -> L_cred = 1):
theorem am_gm (a b : $\mathbb{R}$) (ha : 0 $\le$ a) (hb : 0 $\le$ b) :
    Real.sqrt (a * b) $\le$ (a + b) / 2 := by
  sorry
\end{lean}
Gate~$0\to 1$ checks only syntax and typing. Suppose instead that an agent strengthens the premises to $a \ge 1$ and $b \ge 1$, making the statement strictly weaker because it excludes every pair with $a < 1$ or $b < 1$:
\begin{lean}
-- Weakened L1 declaration (passes G$_{0\to 1}$, fails Gate 1$\to$2: L_tech = 1, cap = 0 -> L_cred = 0):
theorem am_gm_weak (a b : $\mathbb{R}$) (ha : 1 $\le$ a) (hb : 1 $\le$ b) :
    Real.sqrt (a * b) $\le$ (a + b) / 2 := by
  sorry
\end{lean}
Even if \texttt{am\_gm\_weak} were proved without \texttt{sorry} ($L_{\mathrm{tech}} = 3$), restricting the domain fails Gate~$1\to 2$ ($\mathrm{cap} = 0$, so $L_{\mathrm{cred}} = 0$ and $\Delta = +3$). Likewise, replacing $a \ge 0 \land b \ge 0$ by $a + b \ge 0$ alters the domain via Lean's totalised \texttt{Real.sqrt}, which returns~\texttt{0} on negative products (so $\sqrt{10 \times (-1)} = 0 \le 4.5$); although provable by case-splitting and strictly stronger because it also accepts negative-product pairs with $a + b \ge 0$, Gate~$1\to 2$ flags a junk-value caveat because those extra cases rely on \texttt{Real.sqrt} returning~\texttt{0} on negative inputs.

\paragraph{Level 2: Conditional Closure (Staged Blueprint).}
Gate~$1\to 2$ validates the faithful signature (\texttt{ha : 0 $\le$ a}, \texttt{hb : 0 $\le$ b}) and checks that the main proof closes using only audited sub-lemmas. Here, the algebraic square is isolated into an audited $L_1$ stub \texttt{sq\_diff\_nonneg}:
\begin{lean}
-- Audited sub-lemma stub (L_tech = 1, cap = 4 -> L_cred = 1):
lemma sq_diff_nonneg (a b : $\mathbb{R}$) : 0 $\le$ (Real.sqrt a - Real.sqrt b) ^ 2 := by
  sorry

-- Main proof closed locally over the stub (L_tech = 2, L_cred = 2):
theorem am_gm (a b : $\mathbb{R}$) (ha : 0 $\le$ a) (hb : 0 $\le$ b) :
    Real.sqrt (a * b) $\le$ (a + b) / 2 := by
  have h := sq_diff_nonneg a b
  rw [Real.sqrt_mul ha b]
  linarith [Real.sq_sqrt ha, Real.sq_sqrt hb,
    show (Real.sqrt a - Real.sqrt b) ^ 2 =
      Real.sqrt a ^ 2 - 2 * (Real.sqrt a * Real.sqrt b) + Real.sqrt b ^ 2 by ring]
\end{lean}
The body of \texttt{am\_gm} has zero \texttt{sorry} ($L_{\mathrm{tech}}^{\mathrm{loc}} = 3$), yet its dependency \texttt{sq\_diff\_nonneg} is an $L_1$ stub (the minimum technical level over its dependencies is~$1$). By Equations~\eqref{eq:weakest_link}--\eqref{eq:lcred}, $L_{\mathrm{tech}} = 2$ and $L_{\mathrm{cred}} = 2$ ($\Delta = 0$).

\paragraph{Level 3: Isolated Kernel Verification.}
To pass Gate~$2\to 3$, the developer closes the stub in a standalone module (shown below as a self-contained block):
\begin{lean}
lemma sq_diff_nonneg (a b : $\mathbb{R}$) : 0 $\le$ (Real.sqrt a - Real.sqrt b) ^ 2 :=
  sq_nonneg _

theorem am_gm (a b : $\mathbb{R}$) (ha : 0 $\le$ a) (hb : 0 $\le$ b) :
    Real.sqrt (a * b) $\le$ (a + b) / 2 := by
  have h := sq_diff_nonneg a b
  rw [Real.sqrt_mul ha b]
  linarith [Real.sq_sqrt ha, Real.sq_sqrt hb,
    show (Real.sqrt a - Real.sqrt b) ^ 2 =
      Real.sqrt a ^ 2 - 2 * (Real.sqrt a * Real.sqrt b) + Real.sqrt b ^ 2 by ring]

#print axioms am_gm
-- 'am_gm' depends on axioms: [propext, Classical.choice, Quot.sound]
\end{lean}
Every dependency now reaches Level~3, \texttt{\#print axioms} lists only standard axioms, and \texttt{lake build} serialises the \texttt{.olean} artefact. By Equations~\eqref{eq:weakest_link}--\eqref{eq:lcred}, $L_{\mathrm{tech}} = 3$ and $L_{\mathrm{cred}} = 3$ ($\Delta = 0$).

\paragraph{Level 4: Ecosystem Integration.}
Because a standalone proof of \texttt{am\_gm} duplicates algebraic steps already subsumed by \texttt{Mathlib}'s general power-mean theory, Gate~$3\to 4$ refactors it into canonical \texttt{Mathlib} style (with docstrings, implicit arguments, library naming, and zero helper duplication) by deriving the \texttt{$\mathbb{R}$}-typed corollary directly from \texttt{Real.geom\_mean\_le\_arith\_mean2\_weighted} (note that \texttt{Mathlib} already contains \texttt{NNReal.sqrt\_mul\_le\_half\_add} on \texttt{$\mathbb{R}_{\ge 0}$}, verified in snapshot \texttt{0df444a}, whereas no unweighted two-variable \texttt{Real.sqrt} version exists yet):
\begin{lean}
/-- Arithmetic--geometric mean inequality for two non-negative reals. -/
theorem Real.sqrt_mul_le_add_div {a b : $\mathbb{R}$} (ha : 0 $\le$ a) (hb : 0 $\le$ b) :
    Real.sqrt (a * b) $\le$ (a + b) / 2 := by
  have h := @Real.geom_mean_le_arith_mean2_weighted (1 / 2) (1 / 2) a b
    (by norm_num) (by norm_num) ha hb (by norm_num)
  rw [Real.sqrt_mul ha, Real.sqrt_eq_rpow, Real.sqrt_eq_rpow]
  linarith
\end{lean}
Passing Gate~$3\to 4$ then requires passing \texttt{Mathlib}'s linters (\texttt{\#lint}, verified at zero errors across 14~linters on this declaration) and human maintainer Pull Request review (which remains pending here).

\section{Empirical Audit of a Collaborative Formalisation Campaign}
\label{sec:macro_audit}

To test ProofVMM on an active multi-contributor project, we audited the \emph{Beyond $n \log n$} effort \cite{beyondnlogn} and its formalisation missions on the \emph{Prove2Me} platform \cite{prove2me2026} (snapshot: 15~November~2026). The campaign seeks integer multiplication algorithms faster than $O(n \log n)$. Formally, the predicate \texttt{KappaBound($\kappa$)} asserts that there exists a deterministic multitape Turing machine (\texttt{MultitapeTM}) that halts with the correct product for every $n \ge 1$ and achieves runtime $O(n (\lg n)^{1-\kappa})$ for all $n \ge n_0$ (for some threshold $n_0$), where $\lg n = \max(\lceil\log_2 n\rceil, 1)$ and $\kappa > 0$ measures the exponent saving. We checked the shared machine definitions (\texttt{IntMul\_MultitapeModel}) and their plain-English platform descriptions (Appendix~\ref{app:defs}), confirming that they match the standard multitape Turing machine model.

\subsection{Campaign-Wide Audit Results}
\label{sec:macro_scope}

We ran a read-only paginated query across all \textbf{10 missions} in the \emph{Prove2Me} \texttt{integer-\allowbreak multiplication-\allowbreak kappa} campaign.\footnote{All data were collected via read-only \texttt{GET} requests to \texttt{prove2.me/api/v1}; mission roots were identified from module imports and \texttt{/open-leaves}.} Together, these 10 missions list \textbf{415 mission-node entries} (354 \texttt{Proved}, or 85.3\%), corresponding to \textbf{289 entries} (254 \texttt{Proved}, or 87.9\%) after removing the identical 126-entry pool (100~\texttt{Proved}) shared by Rounds~7 and~8 (a small number of forward target entries, such as \texttt{community\_pr62} and \texttt{jain\_round7*}, also appear across predecessor mission listings). Table~\ref{tab:campaign_audit} summarises every mission:

\begin{itemize}
    \item \textbf{Platform status versus credited level.} Five of the 10 missions (50\%) are marked \texttt{Proved} with zero open leaves in their root dependency tree ($L_{\mathrm{tech}} = 3$). After auditing the statement closures of the mission roots and shared base definitions (\texttt{IntMul\_MultitapeModel}), \textbf{3 missions (30\%)}---the three unconditional OpenAI constructions (\texttt{R182}, \texttt{R154}, \texttt{R129} at $\kappa \in \{2^{-182}, 2^{-154}, 2^{-129}\}$)---achieve statement-audited full-scope verification ($L_{\mathrm{cred}} = 3$, $\Delta = 0$). Under Equation~\eqref{eq:lcred}, auditing the root statement closure and base definitions suffices once proof closure passes; we additionally verified that all 21~entries in these three missions (1, 19, and 1~respectively) are \texttt{Proved} with zero custom axioms, \texttt{native\_decide}, or \texttt{unsafe} code in their stored statements and preambles. These credits rest on the platform's server-side attestation of the proofs; independent local replay is future work (Limitation~3).
    \item \textbf{Fixed-constant proofs and shared theorem pools.} Two \texttt{Proved} missions (Round~6 and PR~\#62) drop to $L_{\mathrm{cred}} = 2$ ($\Delta = +1$) because their statements verify fixed numerical grids or finite lookup tables rather than proving the general parametric claim. Although their counts show open entries (\texttt{58/59} and \texttt{73/77}), inspecting those open nodes reveals that they are future target theorems (pointing to PR~\#62 and Round~7) placed in the same workspace folder for navigation; they are not imported by the closed root theorem. Rounds~7 and~8 are faithful \texttt{Open} Level~2 blueprints ($L_{\mathrm{tech}} = L_{\mathrm{cred}} = 2$, $\Delta = 0$) that share the same 126-entry pool and single open sub-goal (\texttt{R8.Bit.Oppbank.nest\_Le\_ND}). Note that a naive single-page API query capped at 100 items missed this open leaf and falsely showed both missions as leaf-free. (Sensitivity check: if fixed-constant scoped proofs were capped at $L_{\mathrm{cred}}=1$ instead of~$2$, Round~6 and PR~\#62 would drop to $L_{\mathrm{cred}}=1$, while the $3/10$ count of $L_{\mathrm{cred}}=3$ missions stays unchanged.)
    \item \textbf{Unsupported stronger targets at the frontier.} The three highest-$\kappa$ missions ($\kappa > 3 \times 10^{-3}$: PR~\#6453, \#7000/\#7001, and PR~\#7113) originate from \emph{conditional} community proposals relying on five unproven upstream assumptions. Yet on \emph{Prove2Me}, each mission specifies an \emph{unconditional} \texttt{KappaBound} target. Proving an unconditional target would establish the conditional proposal, but whilst the target remains open ($L_{\mathrm{tech}} < 3$), the weaker conditional proposal cannot discharge it. Relative to their conditional source provenance, all three are tagged as \emph{unsupported stronger targets} ($L_{\mathrm{cred}} = 0^*$). In \#7000/\#7001, the sole \texttt{Proved} theorem (\texttt{jain\_round6}) applies to a much smaller~$\kappa$ and gives no support to the target.
\end{itemize}

\begin{table}[htbp]
\centering
\scriptsize
\caption{ProofVMM audit of all 10 missions in the \emph{Prove2Me} \texttt{integer-multiplication-kappa} campaign (15 November 2026).}
\label{tab:campaign_audit}
\begin{tabularx}{\textwidth}{@{}l l l c c c c c >{\raggedright\arraybackslash}X@{}}
\toprule
\textbf{Target $\kappa$} & \textbf{Source Claim} & \textbf{Status} & \textbf{Nodes} & \textbf{Leaves} & \textbf{$L_{\mathrm{tech}}$} & \textbf{$L_{\mathrm{cred}}$} & \textbf{$\Delta$} & \textbf{Main Finding} \\
\midrule
$2^{-182}$ & OpenAI R182 \cite{beyondnlogn} & Proved & 1/1 & 0 & 3 & 3 & 0 & Leaf-free; child \texttt{NLogN.KFTop.kappa\_bound} Proved \\
$2^{-154}$ & OpenAI R154 \cite{beyondnlogn} & Proved & 19/19 & 0 & 3 & 3 & 0 & Full machine and parameter bounds Proved \\
$2^{-129}$ & OpenAI R129 \cite{beyondnlogn} & Proved & 1/1 & 0 & 3 & 3 & 0 & Leaf-free; no fixed-grid restrictions \\
$3.67\times 10^{-5}$ & Round~6 \cite{beyondnlogn} & Proved & 58/59 & 0 & 3 & 2 & $+1$ & Scoped R6 arithmetic (\texttt{JainData} / \texttt{CrudeAssembly}) \\
$5.10\times 10^{-5}$ & PR~\#62 \cite{beyondnlogn} & Proved & 73/77 & 0 & 3 & 2 & $+1$ & Scoped instantiation via fixed finite tables (\texttt{J6Seg.*}) \\
$6.40\times 10^{-5}$ & Round~7 \cite{beyondnlogn} & Open & 100/126 & 1 & 2 & 2 & 0 & Shares 126-entry pool with R8; 1 open leaf (\texttt{nest\_Le\_ND}) \\
$1.26\times 10^{-4}$ & Round~8 \cite{beyondnlogn} & Open & 100/126 & 1 & 2 & 2 & 0 & Faithful Level~2 blueprint; 1 open leaf (\texttt{nest\_Le\_ND}) \\
$3.29\times 10^{-3}$ & Cond.\ PR~\#6453 & Open & 1/3 & 1 & 2 & 0* & $+2$ & *Unsupported stronger open target vs.\ 5-assumption PR \\
$3.63\times 10^{-3}$ & Cond.\ \#7000/\#7001 & Open & 1/2 & 1 & 1 & 0* & $+1$ & *Unsupported stronger open target vs.\ conditional PR; unrelated child \\
$3.64\times 10^{-3}$ & Cond.\ PR~\#7113 & Open & 0/1 & 1 & 1 & 0* & $+1$ & *Unsupported stronger open target beyond model bundle \\
\bottomrule
\end{tabularx}
\end{table}

\subsection{Case Study: Interactive Agent Trace on PR~\#6453}
\label{sec:macro_pr6453}

To observe how an AI agent interacts with ProofVMM gates on an open research target, the first author conducted an interactive session (8--9~November~2026) using \texttt{DeepSeek V4.1 Flash} inside SAIR \emph{OpenMath OS} \cite{sair2026openmath} and the \emph{Prove2Me} workspace (Lean~4.33.1, Mathlib commit \texttt{0df444a}). Human prompts steered task selection across turns; full session logs are archived with our scripts. The task targeted \texttt{IntMul.Kappa.conditional\_kappa\_003287014\_pr6453} (\mbox{$\kappa = 1643507/500000000 = 0.003287014$}).

Over the session, the model produced \textbf{21 Lean~4 theorems} accepted on \emph{Prove2Me} (plus 16 definition bundles) and \textbf{1 local check} (\texttt{replay\_holds}, making 22 verified artefacts in total; exact details and equations appear in Appendix~\ref{app:pr6453}). All 21 platform theorems compile with zero \texttt{sorry} and standard axioms only:
\begin{itemize}
    \item \emph{Cost and Contraction Bounds (8 theorems):} verify numerical inequalities showing that recursive multiplication costs shrink geometrically across recursion levels, requiring 11-decimal Taylor enclosures in Lean.
    \item \emph{Program Execution Rules (7 theorems):} verify structural invariants of PR~\#6453's 17,926-step schedule over 1,095 columns (showing that execution legality depends only on column state labels and that data updates are linear).
    \item \emph{Modular Algebra (6 theorems):} prove abstract polynomial and Chinese Remainder Theorem (CRT) matrix identities used by the algorithm's algebraic layer.
\end{itemize}

\begin{figure}[htbp]
\centering
\begin{tikzpicture}[>=Stealth, every node/.style={rounded corners=2pt, align=center, font=\scriptsize, inner sep=5pt}]
  \node[draw=dvorakRed, thick, fill=red!6, minimum width=5.6cm, minimum height=1.0cm] (root) at (0, 3.1)
    {\textbf{\texttt{conditional\_kappa\_003287014\_pr6453}} (\texttt{Open})\\
     $L_{\mathrm{tech}}=2$ (sketch accepted),\quad $\mathbf{L_{\mathrm{cred}}=0^*}\;(\Delta=+2)$};

  \node[draw=teal!70!black, fill=teal!6, minimum width=4.0cm, minimum height=0.95cm] (glue) at (-3.6, 1.4)
    {\texttt{kappaBound\_of\_polylog\_slack}\\
     $L_{\mathrm{tech}}=3,\; L_{\mathrm{cred}}=3$};

  \node[draw=dvorakRed, thick, fill=red!6, minimum width=5.0cm, minimum height=0.95cm] (leaf) at (2.4, 1.4)
    {\textbf{\texttt{pr6453\_machine}} (\texttt{Open}, \texttt{sorry})\\
     $L_{\mathrm{tech}}=1,\; \mathbf{L_{\mathrm{cred}}=0^*}$ (unsupported target)};

  \draw[->, thick, teal!70!black] (glue) -- (root);
  \draw[->, thick, dvorakRed] (leaf) -- (root);

  \node[draw=teal!70!black, fill=teal!5, minimum width=3.4cm, minimum height=1.15cm] (c1) at (-4.9, -0.65)
    {\textbf{Cost Bounds (8)}\\
     $L_{\mathrm{tech}}=3,\; L_{\mathrm{cred}}=3^\dagger$\\
     \emph{e.g.}\ \texttt{moment\_kappa\_gt}};

  \node[draw=teal!70!black, fill=teal!5, minimum width=3.8cm, minimum height=1.15cm] (c2) at (0.0, -0.65)
    {\textbf{Program Replay (7 + 1)}\\
     $L_{\mathrm{tech}}=3^\dagger$ (7 thms)\\
     \texttt{replay\_holds}: \texttt{.olean} OOM ($L_2$)};

  \node[draw=orange!80!black, fill=orange!6, minimum width=3.2cm, minimum height=1.15cm] (c3) at (4.9, -0.65)
    {\textbf{Modular Algebra (6)}\\
     $L_{\mathrm{tech}}=3$, $L_{\mathrm{cred}}=3^\dagger / 0^\dagger$\\
     $1152^2$ Mis-indexing};

  \node[draw=dvorakGray, dashed, inner sep=7pt, fit=(c1)(c2)(c3),
        label={[font=\scriptsize\itshape, text=dvorakGray]below:{21 Proved Theorems + 16 Def.\ Bundles: off critical path (none closes \texttt{pr6453\_machine})}}] (box) {};

  \draw[dashed, dvorakRed] (box.north -| leaf.south) -- (leaf.south)
    node[midway, fill=white, font=\scriptsize\bfseries, text=dvorakRed] {$\times$ not ancestors};
\end{tikzpicture}
\caption{ProofVMM dependency view of PR~\#6453 on \emph{Prove2Me}. Despite 21 kernel-checked auxiliary theorems (and 1 local RAM-checked check), all auxiliary proofs lie off the critical path ($\times$), leaving the unsupported target at \texttt{pr6453\_machine} open ($L_{\mathrm{cred}}=0^*$).}
\label{fig:pr6453_dag}
\end{figure}
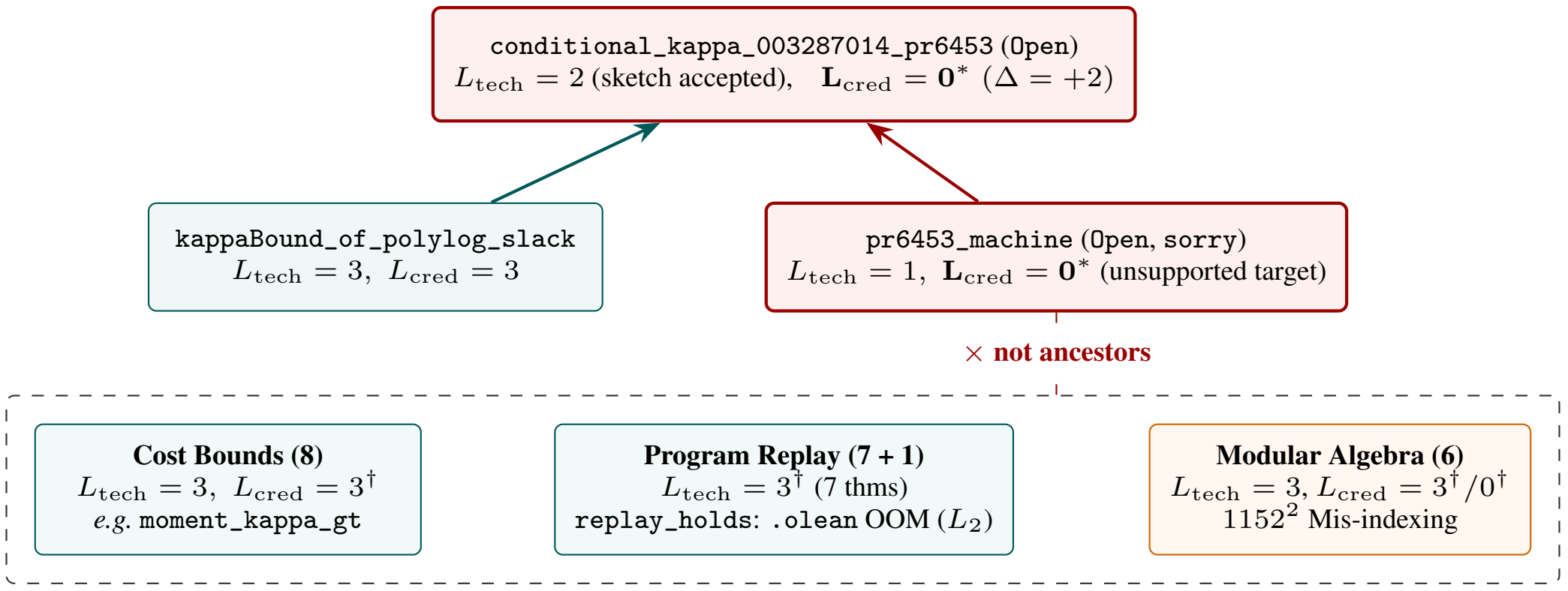

We call a theorem on a mission's \emph{critical path} if it is an ancestor of at least one open leaf in the dependency graph (so proving it helps close that leaf). In our PR~\#6453 trace (Figure~\ref{fig:pr6453_dag} and Table~\ref{tab:pr6453_audit}), \textbf{21 of 21 (100\%)} locally proved theorems lie off the critical path: they prove finite certificate numbers or general algebraic facts without constructing the Turing machine or resolving the five upstream assumptions. Similarly, in mission \#7000/\#7001, the sole \texttt{Proved} node lies off the critical path ($1/1$). Separately, the 17,926-step local schedule check (\texttt{replay\_holds}, our 22nd item) passed Lean's kernel in memory in 2m\,32s without \texttt{native\_decide}, yet exhausted the OpenMath~OS cloud container's RAM when saving its \texttt{.olean} file, remaining at~$L_2$ under our resource-budget criterion.

\begin{table}[htbp]
\centering
\footnotesize
\caption{ProofVMM breakdown of the 22 PR~\#6453 trace artefacts (compared with the \texttt{NLogN.FIN} baseline). Here $*$ marks unsupported stronger open targets relative to the conditional PR, whereas $\dagger$ denotes standalone finite or abstract statements evaluated on their own formal claims (off the mission's critical path).}
\label{tab:pr6453_audit}
\begin{tabularx}{\textwidth}{@{}lccc>{\raggedright\arraybackslash}X@{}}
\toprule
\textbf{Artefact / Node Group} & \textbf{$L_{\mathrm{tech}}$} & \textbf{$L_{\mathrm{cred}}$} & \textbf{$\Delta$} & \textbf{ProofVMM Diagnosis} \\
\midrule
\texttt{NLogN.FIN.theorem\_1\_1\_nlogn} & 3 & 3 & 0 & Baseline: 9 proved children, 0 open leaves; full $L_3$ machine proof \\
\texttt{conditional\_kappa\_003287014\_pr6453} & 2 & 0* & $+2$ & Compiles as sketch, *unsupported through \texttt{pr6453\_machine} \\
\texttt{pr6453\_machine} & 1 & 0* & $+1$ & *Unsupported stronger target: unconditional type vs.\ 5 assumptions \\
\texttt{Pr6453.*} cost \& replay (15 thms) & 3 & 3$^\dagger$ & 0 & $^\dagger$Locally $L_3$ for finite PR data/rules; off critical path (15/15) \\
\texttt{CRTBasis}/\texttt{Pr6453Encoding}/\texttt{Root}/\texttt{Site} (4 thms) & 3 & 3$^\dagger$ & 0 & $^\dagger$Valid abstract/data $L_3$ lemmas; off critical path (4/4) \\
\texttt{Pr6453Incidence}/\texttt{Pr6453Tensor} (2 thms) & 3 & 0$^\dagger$ & $+3$ & $^\dagger$Valid abstract lemmas, mis-indexed ($1152^2$) for PR~\#6453 \\
\texttt{replay\_holds} (22nd item, local) & 2 & 2 & 0 & Passes kernel in RAM (2m\,32s), fails \texttt{.olean} export (OOM) at $G_{2\to 3}$ \\
\bottomrule
\end{tabularx}
\end{table}

\subsection{Taxonomy of Error Modes Observed in the Audit}
\label{sec:vmm_taxonomy}

Across the campaign audit and the PR~\#6453 trace, we observed three distinct failure modes (note that \emph{Scope Mismatch} refers to a structural mismatch between the platform target and the PR's stated conditions, not contributor intent):

\begin{enumerate}
    \item \textbf{Silent Totalisation and Elaboration Drift.} Lean~4 defines total functions that return default values outside their usual mathematical domain without raising compiler errors. Examples include \texttt{sInf $\emptyset$ = 0}, division by zero (\texttt{x / 0 = 0}), divergent infinite sums (\texttt{tsum} returning~\texttt{0}), and writing \texttt{(12 : $\mathbb{R}$) \^{} (1 / 7)}. The last expression occurred during PR~\#6453, where \texttt{1 / 7} elaborated as natural-number division (\texttt{0}) and silently evaluated $12^{1/7}$ to~$1$:
\begin{lean}
-- Compiles via Mathlib junk values (0 / 0 = 0 and log 0 = 0):
example (x : $\mathbb{R}$) : Real.log (x / x) = 0 := Real.log_div_self x

-- Compiles because 1 / 7 elaborates in $\mathbb{N}$ to 0, giving $12^0 = 1$:
example : (12 : $\mathbb{R}$) ^ (1 / 7) = 1 := by norm_num
\end{lean}
    \item \textbf{Specification Alteration (Weakening and Scope Mismatch).} Formalisations either weaken a source claim (adding extra hypotheses or narrowing conclusions, giving $L_{\mathrm{cred}} = 0$) or (as in PR~\#6453, \#7000/\#7001, and PR~\#7113) bind a conditional source construction (relying on five unproven upstream assumptions) to an unconditional open target signature (\texttt{KappaBound}), leaving the target unsupported by the source (Appendix~\ref{app:pr6453}).
    \item \textbf{Mis-indexed Specifications and Off-Path Theorems.} An agent can produce valid $L_3$ proofs for the wrong data layout. In PR~\#6453, the agent spent several turns proving lemmas for a $1152^2$ input-vector product (\texttt{Pr6453Incidence.spec\_of\_basis} and \texttt{Pr6453Tensor.checkBasis\_sound}) before checking the project's \texttt{ROOT.json} file, where $1152 = 2 \times 16 \times 36$ actually counts coefficients of an internal address polynomial rather than input vectors. Similarly, an agent can accumulate proved lemmas outside the critical path of any open leaf ($21/21$ on PR~\#6453). Meanwhile, external Python checks (such as 60-digit floating-point bounds and trace replays) remain non-kernel evidence under Convention~(\ref{conv:external}).
\end{enumerate}

\paragraph{Summary.} Overall, raw counts of \texttt{Proved} nodes do not distinguish finished proofs from fixed-constant shortcuts, mis-indexed lemmas, or unsupported targets. Applying ProofVMM separates proof closure ($L_{\mathrm{tech}}$) from audited statement faithfulness ($L_{\mathrm{cred}}$).

\section{Related Work}
\label{sec:related}

Recent research on AI-assisted formalisation in Lean~4 falls into three groups; to our knowledge, none combines a five-tier artefact lifecycle model with statement-versus-proof auditing of a live multi-mission platform.

\paragraph{Compile--Faithfulness, Statement Audits, and Verification Criteria.}
Several studies show that compiling Lean~4 code does not guarantee mathematical faithfulness. \emph{Beyond Compilation} \cite{zhang2026compilationevaluatingfaithfulnaturallanguagetolean}, \emph{BEq+} \cite{poiroux2025reliableevaluationbenchmarksstatement}, and Mohammad and Sheikh \cite{mohammad2026the} measure statement-level equivalence and faithfulness gaps, \emph{claimcheck} \cite{amster2026claimcheck} detects specification drift via back-translation to natural language, and \emph{SemantEq} \cite{li2026semanteqfaithfulbidirectionalequivalence} evaluates bidirectional equivalence between formal statements. Allaham et~al.\ \cite{allaham2026elicitingspecificationsdoesinteractive} show that informal problem statements often omit edge cases needed by the verifier during $L_0 \to L_1$ translation. In PDE analysis, Bastounis et~al.\ \cite{bastounis2026navierstokeslosttranslationlean} show that an autoformalised Navier--Stokes proof alters the paper's derivative hypotheses, proving via Hilbert's Tenth Problem that checking semantic preservation is undecidable and that resolving informal ambiguities can lie arbitrarily high in the Solvability Complexity Index (SCI) hierarchy. For applied mathematics, \emph{Construction-Verification} \cite{yang2026constructionverificationbenchmarkappliedmathematics} documents non-constructive existence shortcuts in algorithmic proofs. In parallel, Buzzard's analysis of what makes a Lean proof valid \cite{Buz26} synthesises the consensus criteria for headline machine-checked proofs: matching a trusted canonical statement, a clean build, verifying that the target theorem is strictly what was proved, and passing independent kernel verification (e.g.\ via \texttt{leanchecker} or \texttt{nanoda\_lib}). Closest to our work, \emph{LeanTrail} \cite{venturim2026leantrail} aligns structured \LaTeX\ papers to single Lean repositories via LLM statement judging and deterministic \texttt{sorryAx} checks, whilst \emph{Lean Atlas} \cite{yanahama2026leanatlasintegratedproof} organises human semantic review over Lean dependency graphs. However, \emph{LeanTrail} addresses single paper--repository pairs, tracks only \texttt{sorryAx} (noting project axioms in certificates rather than grading them), omits \texttt{Mathlib} integration ($L_4$), and merges weakened statements into a single \emph{Partial} state rather than separating $L_{\mathrm{tech}}$ from $L_{\mathrm{cred}}$.

\paragraph{Multi-Level Coverage Libraries and Foundational Kernels.}
Benchmarks such as \emph{SorryDB} \cite{letson2026sorrydbaiproverscomplete} and \emph{AnnalsChallenge} \cite{browning2026datasetformalisedtheoremstatements} evaluate provers on repository-scale codebases and curated theorems. Two projects use multi-level tracking: \emph{GenLimitLib} \cite{li2026genlimitlibformallibrarylanguage} reports per-claim coverage across three audit tiers in a 30-paper learning-theory library, and \emph{Lean Blueprint} \cite{massot_leanblueprint} visualises proof status across dependency graphs. Neither defines formal transition gates with scoped-claim caps or cross-mission deficits. From a foundational angle, \emph{Eigenius} \cite{will2026eigeniustypedknowledgegraphdbms} designs a new kernel that retains distinct proof terms and tracks external assumptions (\emph{Adopt}, \emph{Discharge}, \texttt{Relativize}); by contrast, ProofVMM audits unmodified Lean~4 and \texttt{Mathlib} codebases.

\paragraph{Collaborative Formalisation Platforms.}
Collaborative platforms such as \emph{Beyond $n \log n$} \cite{beyondnlogn}, \emph{Prove2Me} \cite{prove2me2026}, and SAIR \emph{OpenMath OS} \cite{sair2026openmath} coordinate human and AI contributors via dependency graphs, accepted proof sketches, and plain-English statement summaries. ProofVMM complements these platforms by propagating proof closure and auditing statement fidelity, exposing fixed-constant shortcuts, off-path lemmas, and scope mismatches that remain invisible under \texttt{Open} versus \texttt{Proved} counters.

\section{Discussion, Limitations, and Conclusion}
\label{sec:discussion}

\paragraph{Practical Recommendations.}
Our audit suggests three simple improvements for collaborative platforms and agent harnesses: (1)~show statement-audited $L_{\mathrm{cred}}$ scores alongside raw \texttt{Proved} counts so disconnected lemmas do not inflate progress; (2)~keep conditional hypotheses explicitly inside mission signatures rather than posting unconditional targets for conditional proposals; and (3)~flag high-risk Lean patterns (such as natural-number division in real exponents, \texttt{sInf $\emptyset$}, and \texttt{native\_decide}) automatically before human Gate~$1\to 2$ review.

\paragraph{Scalability and Headline Open Conjectures.}
These findings also illuminate what is required as AI provers scale to headline open problems. Whereas an autoformalised Navier--Stokes claim revealed silent derivative-hypothesis drift upon audit \cite{bastounis2026navierstokeslosttranslationlean}, the November~2026 Lean~4 proof of the Riemann Hypothesis announced by Alp\"oge ($\sim$17.4~million lines of Lean code, developed with Claude) demonstrates how statement-closure auditing operates at massive scale \cite{LFM26,Sut26}: rather than manually reading every intermediate lemma, community and Lean~FRO (Focused Research Organization) verification focused on confirming that the formal statement and unfolded definitions match \texttt{Mathlib}'s \texttt{RiemannHypothesis} (using \texttt{comparator} and \texttt{con-leche --compare}) whilst running independent kernel checkers (\texttt{leanchecker} and \texttt{nanoda\_lib}) over the full proof term \cite{LFM26,Sut26}. Under Equations~\eqref{eq:weakest_link}--\eqref{eq:lcred}, such an archive would be graded as $L_{\mathrm{tech}} = 3$ and $L_{\mathrm{cred}} = 3$, though it has not yet been refactored or reviewed for \texttt{Mathlib} integration ($L_4$). Separating statement-closure auditing from proof-closure checking is therefore essential if verification pipelines are to scale alongside multi-million-line AI proofs.

\paragraph{Limitations and Threats to Validity.}
This study has five main limitations. First, our audit examines one collaborative ecosystem (\emph{Beyond $n \log n$} / \emph{Prove2Me}) and one interactive agent trace (\texttt{DeepSeek V4.1 Flash} on PR~\#6453); testing additional frontier models, computing error frequencies across all 289 entries, validating ProofVMM scores against independent ground truth such as \emph{LeanTrail}'s 98-item benchmark \cite{venturim2026leantrail}, and auditing a second corpus remain future work. Second, semantic auditing at Gate~$1\to 2$ was conducted by a single auditor using Appendix~\ref{app:checklist}; dual-annotator review with inter-rater agreement (Cohen's~$\kappa$) is needed to measure consistency. Third, whilst we inspected mission roots, statement summaries, \texttt{IntMul\_MultitapeModel}, and the stored statements and preambles of all 21~entries in the three $L_{\mathrm{cred}} = 3$ missions, third-party proof bodies across the 289~entries are not exposed in bulk by the API and rely on platform server-side verification for internal \texttt{\#print axioms} cleanliness (independent local re-compilation of all third-party mission bodies remains future work). Fourth, because no audited mission satisfied Gate~$3\to 4$, Level~4 is demonstrated only through the refactoring and \texttt{\#lint} stage of our AM--GM worked example (maintainer review remains pending). Fifth, API limits required reconstructing roots from module imports, and \texttt{.olean} memory limits reflect the OpenMath~OS container budget.

\paragraph{Ethics, Artefacts, and AI Use.}
All platform data were collected strictly via read-only \texttt{GET} queries (15~November~2026 snapshot). All audit scripts, raw JSON snapshots (stripped of API credentials), verified Lean~4.33.1 snippet checks (compiled against \texttt{Mathlib} commit \texttt{0df444a}), the audit checklist, and session logs will be archived publicly at \url{https://github.com/gustason/vmm-audit}. AI tools (\texttt{DeepSeek V4.1 Flash} inside OpenMath~OS) served as the experimental subject on PR~\#6453 and \texttt{Gemini 4 Hydrogen} assisted with language editing; all ProofVMM definitions, manual audits, and final text were verified by the authors.

\paragraph{Conclusion.}
As AI systems become adept at writing code that passes Lean's type checker, the main bottleneck shifts from syntax to semantic faithfulness and dependency structure. ProofVMM replaces binary ``verified'' labels with five inspectable tiers, four transition gates, and a clean separation between proof closure and statement-closure credit. Applied to a live collaborative campaign, ProofVMM exposes how fixed-constant shortcuts, disconnected lemmas, and scope mismatches separate apparent compilation from genuine theorem closure.

\bibliographystyle{alpha}
\bibliography{references}

\appendix
\section{PR~\#6453 Interactive Trace Inventory and Code Excerpt}
\label{app:pr6453}

\paragraph{Mathematical summary of PR~\#6453.}
Sub-$n\log n$ multiplication schemes divide an $n$-bit product recursively into smaller sub-calls indexed by rank~$r$, tracked by counts of productive targets~$T_r$ and paid calls~$P_r$. The $n(\lg n)^{1-\kappa}$ exponent survives recursion if the rank-moment inequality $\sum_r T_r r^{1-\kappa} > \sum_r P_r r^{1-\kappa}$ holds, ensuring per-level contraction $\gamma = (\sum_r P_r r^{1-\kappa})/(\sum_r T_r r^{1-\kappa}) < 1$ and total cost bounded by $1/(1-\gamma)$ times the root cost. At $\kappa = 0.003287014$, $1-\gamma \approx 6.9 \times 10^{-9}$ ($1/(1-\gamma) \approx 1.45 \times 10^8$), requiring 11-decimal Taylor bounds in Lean. Combinatorially, the PR encodes a 17,926-step program over 1,095 columns (799 written) intended to evaluate a Chinese Remainder Theorem (CRT) module identity, whilst leaving the \texttt{MultitapeTM} machine, caller correctness, and five inherited assumptions (\texttt{native\_all\_columns}, \texttt{owned\_erasure}, \texttt{compiler}, \texttt{precision\_tape}, \texttt{analytic\_recovery}) unproved.

\paragraph{Itemised 21 platform theorems (+ 1 local check, Lean~4.33.1, Mathlib \texttt{0df444a}).}
\emph{Cost and Contraction (8):} \texttt{Pr6453.moment\_kappa\_gt}, \texttt{Pr6453.invoice}, \texttt{Pr6453.outer\_ledger}, \texttt{Pr6453.recursion}, \texttt{Pr6453.strictification}, \texttt{Pr6453.shifted\_lower}, \texttt{Pr6453.shifted\_upper}, \texttt{Pr6453.tail\_ratios}.
\emph{Program Replay (7 + 1):} \texttt{Pr6453.word\_receipts}, \texttt{Pr6453.native\_helpers}, \texttt{Pr6453.replay\_composes}, \texttt{Pr6453.replay\_frame\_determined}, \texttt{Pr6453.replay\_const}, \texttt{Pr6453.replay\_track}, \texttt{Pr6453.replay\_linear}, plus the 22nd item \texttt{Pr6453.replay\_holds} (17,926-step frame check verified in RAM in 2m\,32s, failing \texttt{.olean} export by OOM).
\emph{Modular Algebra (6):} \texttt{CRTBasis.crt\_basis\_mul}, \texttt{Pr6453Encoding.crt\_encoding}, \texttt{Pr6453Incidence.spec\_of\_basis}, \texttt{Pr6453Tensor.checkBasis\_sound}, \texttt{Pr6453Root.address\_data}, \texttt{Pr6453Site.site\_identities}.
(All 21 platform theorems compile with 0~\texttt{sorry} and axioms \texttt{[propext, Classical.choice, Quot.sound]}, alongside 16 definition bundles; none discharges \texttt{pr6453\_machine}.)

\begin{lean}
-- Excerpt requiring project-local IntMul definitions (MultitapeTM, KappaBound, lg):
-- Open leaf on Prove2Me (L_tech = 1, L_cred = 0*: unsupported target vs conditional PR):
theorem IntMul.Kappa.pr6453_machine :
    $\exists$ (M : MultitapeTM) (g C : $\mathbb{R}$) (A n$_0$ : $\mathbb{N}$),
      0.003287014 < g $\land$ 0 < C $\land$
      ($\forall$ n : $\mathbb{N}$, 1 $\le$ n $\to$ $\exists$ T : $\mathbb{R}$, MultipliesAt M n T) $\land$
      ($\forall$ n : $\mathbb{N}$, n$_0$ $\le$ n $\to$
        MultipliesAt M n (C * n * (lg n : $\mathbb{R}$) ^ (1 - g) * (Real.log (lg n) + 1) ^ A)) := by
  sorry

-- Accepted sketch reduction (L_tech = 2, L_cred = 0* via Eq. (3) since cap(pr6453_machine) = 0):
theorem solution : IntMul.KappaBound (0.003287014 : $\mathbb{R}$) := by
  obtain $\langle$M, g, C, A, n$_0$, hg, _hC, hcorr, hT$\rangle$ := IntMul.Kappa.pr6453_machine
  exact IntMul.Kappa.kappaBound_of_polylog_slack 0.003287014 g C A hg M hcorr n$_0$ hT
\end{lean}

\section{Audit of Shared Base Definitions (\texttt{IntMul\_MultitapeModel})}
\label{app:defs}

To verify that statement-audited Level~3 ratings rest on faithful machine definitions, we inspected the platform's \texttt{IntMul\_MultitapeModel} module and the independent plain-English statement summaries across the audited missions:
(1)~\emph{Tapes and alphabet:} deterministic Turing machine over a finite alphabet \texttt{Sym} containing five pairwise-distinct symbols (blank, start marker, \texttt{0}, \texttt{1}, \texttt{\#}), finitely many states with distinct start and halting states (frozen upon halting), and $k \ge 2$ one-way-infinite tapes (tape~0 read-only input, tape~1 output);
(2)~\emph{Complexity predicate:} \texttt{MultipliesAt M n T} requires correct multiplication for all $n$-bit inputs ($n \ge 1$) within $T$ steps, and \texttt{KappaBound $\kappa$} asserts two conjuncts: correctness for all input lengths ($\forall n \ge 1,\ \exists T,\ \texttt{MultipliesAt}\ M\ n\ T$) and an eventual runtime bound ($\exists c > 0,\ \exists n_0,\ \forall n \ge n_0,\ 1 \le n \to \texttt{MultipliesAt}\ M\ n\ (c \cdot n \cdot (\lg n)^{1-\kappa})$), where $\lg n = \max(\lceil\log_2 n\rceil, 1)$ (the polylogarithmic-slack variant used in \texttt{pr6453\_machine} multiplies by $(\ln(\lg n)+1)^A$);
(3)~\emph{Register-machine bridge:} \texttt{NLogN.FIN} reduces verified register-machine bounds (\texttt{RM}/\texttt{StepOK}) into \texttt{IntMul.MulTimeBound} over the same multitape model.

\section{Gate~$1\to 2$ Semantic Audit Checklist}
\label{app:checklist}

Each audited statement was checked against five criteria:
(1)~every hypothesis in the source claim matches a Lean hypothesis (noting omissions or added conditions);
(2)~conclusion strength matches the source (no dropped bounds or unproved terms);
(3)~quantifier order and parameter dependencies are preserved;
(4)~default Mathlib totalisations (\texttt{sInf $\emptyset$}, \texttt{x / 0}, \texttt{Real.sqrt} on negatives, \texttt{gradient} at non-differentiable points, \texttt{tsum} on divergent series, \texttt{$\mathbb{N}$} division/subtraction) do not alter the claim;
(5)~explicit non-vacuity and provenance boundaries (inherited conditional assumptions) are preserved.

\end{document}